\documentclass[twocolumn]{aastex631}

\usepackage{ulem}
\usepackage{xcolor}
\usepackage{graphicx}   
\usepackage{amsmath}    
\usepackage{amssymb}    
\usepackage{diagbox}    
\usepackage{subfigure}
\usepackage{float}
\usepackage{booktabs}
\usepackage{threeparttable}
\usepackage{color}
\usepackage{subfigure}
\usepackage{cancel}
\usepackage{soul}
\usepackage{graphicx}
\usepackage{float}
\usepackage{natbib}
\usepackage{dcolumn}
\usepackage{bm}
\usepackage{ulem}
\usepackage{color}
\usepackage{txfonts}
\usepackage{rotating}
\usepackage{booktabs}
\usepackage{array}
\usepackage{tabularx}
\usepackage{threeparttable}
\usepackage{diagbox}
\usepackage{multirow}
\usepackage{makecell}
\usepackage{longtable}

\shorttitle{A possible formation scenario of the $Gaia$ BH1: inner binary merger in triple systems}
\shortauthors{Zhuowen Li et al.}

\begin{document}

\title{A possible formation scenario of the $Gaia$ BH1: inner binary merger in triple systems}

\correspondingauthor{Chunhua Zhu, Guoliang L\"{u}}
\email{chunhuazhu@sina.cn, guolianglv@xao.ac.cn}

\author[0009-0006-1716-357X]{Zhuowen Li}
\email{ZhuoWenli2024@163.com}
\affiliation{School of Physical Science and Technology, Xinjiang University, Urumqi, 830046, China}

\author{Chunhua Zhu}
\affiliation{School of Physical Science and Technology, Xinjiang University, Urumqi, 830046, China}

\author[0000-0002-3849-8962]{Xizhen Lu}
\affiliation{School of Physical Science and Technology, Xinjiang University, Urumqi, 830046, China}

\author[0000-0002-3839-4864]{Guoliang L\"{u}}
\affil{Xinjiang Observatory,
the Chinese Academy of Sciences, Urumqi, 830011, China}
\affil{School of Physical Science and Technology,
Xinjiang University, Urumqi, 830046, China}

\author{Lin Li}
\affiliation{School of Physical Science and Technology, Xinjiang University, Urumqi, 830046, China}

\author{Helei Liu}
\affiliation{School of Physical Science and Technology, Xinjiang University, Urumqi, 830046, China}

\author{Sufen Guo}
\affiliation{School of Physical Science and Technology, Xinjiang University, Urumqi, 830046, China}

\author[0000-0001-8493-5206]{Jinlong Yu}
\affiliation{College of Mechanical and Electronic Engineering, Tarim University, Alar, 843300, China}

\begin{abstract}
Based on astrometric measurements and spectral analysis from $Gaia$ DR3, two quiescent black hole (BH) binaries, $Gaia$ BH1 and BH2, have been identified. Their origins remain controversial, particularly for $Gaia$ BH1. By considering a rapidly rotating ($\omega/\omega_{\rm crit} = 0.8$) and strongly magnetized ($B_{\rm 0} = 5000$ G) merger product, we find that, at typical Galactic metallicity, the merger product can undergo efficient chemically homogeneous evolution (CHE). This results in the merger product having a significantly smaller radius during its evolution compared to that of a normally evolving massive star. Under the condition that the initial triple stability is satisfied, we use the Multiple Stellar Evolution (MSE) code and the MESA code to identify an initial hierarchical triple that can evolve into $Gaia$ BH1. It initially consists of three stars with masses of 9.03 $M_{\odot}$, 3.12 $M_{\odot}$, and 1 $M_{\odot}$, with inner and outer orbital periods of 2.21 days and 121.92 days, and inner and outer eccentricities of 0.41 and 0.45, respectively. This triple initially experiences triple evolution dynamics instability (TEDI) followed by Roche lobe overflow (RLOF). During RLOF, the inner orbit shrinks, and tidal effects gradually suppress the TEDI. Eventually, the inner binary undergoes a merger through contact (or collision). Finally, using models of rapidly rotating and strongly magnetic stars, along with standard core-collapse supernova (SN) or failed supernova (FSN) models, we find that a PMB consisting of an 12.11 $M_{\odot}$ merger product and a 1 $M_{\odot}$ companion star (originally an outer tertiary) can avoid RLOF. After a SN or FSN with a low ejected mass of $\sim$0.22 $M_{\odot}$ and a low kick velocity ($46^{+25}_{-33}$ ${\rm km/s}$ or $9^{+16}_{-8}$ ${\rm km/s}$), the PMB can form $Gaia$ BH1 in the Galactic disk.
\end{abstract}

\keywords{ stars: black holes-stars:evolution}

\section{Introduction}
Black hole (BH) binaries are the key objects for understanding BH nature and binary evolution. Among the most studied BH binaries are BH X-ray binaries, which are detected as a result of X-ray radiation caused by BH accreted matter \citep{2006ARA&A..44...49R,2014SSRv..183..223C}. However, as early as the 1960s, \cite{1966SvA....10..251G} predicted another type of BH binary that does not emit X-rays. These binaries are detected through astrometric measurements and/ or spectral analysis.

Recently, in $Gaia$ DR3 and the pre-release of $Gaia$ DR4, three quiescent BH binaries $-$ $Gaia$ BH1, BH2, and BH3 $-$ were discovered by combining precise radial velocity measurements with $Gaia$ astrometry \citep{2023MNRAS.518.1057E,2023AJ....166....6C,2023ApJ...946...79T,2023MNRAS.521.4323E,2024A&A...686L...2G}. \cite{2017ApJ...850L..13B} and \cite{2022ApJ...931..107C} have also successively estimated that $Gaia$ can detect 3800$\sim$12000 and 30$\sim$300 BHs, respectively. Based on very close distances of $Gaia$ BH1, BH2 and BH3 which are only a few hundred to over a thousand pc away from the solar system, \cite{2024PASP..136b4203R} estimated that non-accreting BH binaries may be more common than BH X-ray binaries. Compared to the known BH X-ray binaries, the orbital periods of these quiescent binaries are much longer (For $Gaia$ BH1 and BH2 they are $\sim$186 days and $\sim$1277 days, respectively \citep{2023MNRAS.518.1057E,2023MNRAS.521.4323E}). \cite{2023MNRAS.518.1057E} discussed the binary origin of $Gaia$ BH1. In standard binary evolution, the progenitor of $Gaia$ BH1 must undergo a common envelope evolution (CEE). Simulating realistic CEE is very challenging \citep{2013A&ARv..21...59I}. In popular models, CEE mainly depends on the combined parameter $\alpha_{\rm CE}\lambda$, where $\alpha_{\rm CE}$ is the efficiency of the orbital energy used to expel the envelope of the donor, and $\lambda$ parameterizes the envelope structure of the donor. Typically, $\alpha_{\rm CE} \le 1$, and $\lambda \sim 0.5$. Although $\lambda$ may reach up to several or even several hundreds for a low or medium-mass star, it is smaller than 1 for a star with a mass higher than 15 $M_{\odot}$ \citep{2010ApJ...716..114X}. In order to avoid binary merger during CEE, \cite{2023MNRAS.521.4323E} considered that $Gaia$ BH1 can be formed via standard binary evolution only if $\alpha_{\rm CE} \ge 5$. Very recently, \cite{2024arXiv240313579K} successfully simulated the formation of $Gaia$ BH1 and BH2 via binary evolution with $\alpha_{\rm CE} = 1$ and $\lambda$ is calculated using the fitting formula from \cite{2010ApJ...716..114X}. In their models, massive star must lose $\sim 50\%$ of its initial mass when it evolves into the core-helium burning giant. When CEE occurs, the envelope mass of the massive star is $\sim 3 M_{\odot}$, which is much lower than $\sim 20 M_{\odot}$ in \cite{2023MNRAS.518.1057E}.

On the other hand, \cite{2023MNRAS.518.1057E}, \cite{2023MNRAS.521.4323E}, \cite{2023HEAD...2011644D}, \cite{2024ApJ...964...83G}, \cite{2023MNRAS.526..740R}, and \cite{2024MNRAS.527.4031T} also discussed the formation scenarios involving triple systems for the $Gaia$ BH1. They considered that $Gaia$ BH1 may originate from triple systems or dynamical exchanges in an open cluster. In the evolution simulations in hierarchical triple systems by \cite{2024ApJ...964...83G},
the von Zeipel-Lidov-Kozai (ZLK) effect \citep{1910AN....183..345V,1962P&SS....9..719L,1962AJ.....67..591K,2016ARA&A..54..441N} can delay the onset of the inner binary CEE in a triple system and allow for more CEE to occur at wider orbits. It means that there is more orbital energy available to eject the CE. \cite{2023MNRAS.526..740R}, \cite{2023HEAD...2011644D} and \cite{2024MNRAS.527.4031T} suggested that $Gaia$ BH1 could form through dynamical interactions in a young star cluster or through dynamical capture in an open star cluster although it is in the Galactic field, respectively.

Therefore, the origin of $Gaia$ BH1 is still debated. As discussed by \cite{2023MNRAS.521.4323E}, the progenitor of the $Gaia$ BH1 might originate from the evolution of the merger product of an inner binary in a hierarchical triple system. Unlike the simulations of hierarchical triple systems by \cite{2024ApJ...964...83G}, the products of inner binary mergers may follow a different evolutionary path compared to normal stars \citep{2014ApJ...796..121J,2023MNRAS.521.4323E}. If the merger product undergoes chemical homogeneous evolution (CHE) triggered by rapid rotation and evolves directly into a helium-rich star, it can almost avoid filling its Roche lobe \citep{2012RvMP...84...25M,2015A&A...581A..15S,2016MNRAS.458.2634M,2016A&A...588A..50M,2016MNRAS.460.3545D,2020MNRAS.499.5941D,2022PhR...955....1M}, allowing this Sun-like star (originally an outer tertiary) to survive. Eventually, the merger product evolves into a BH through a supernova (SN) or failed supernova (FSN) \citep{2011ApJ...730...70O,2023ApJ...952...79L}.

In observational statistics, approximately (35$\pm$3)\% of massive stars are born in triple systems and (38$\pm$11)\% are born in quadruple systems \citep{2017ApJS..230...15M}. This means that massive stars are very common in multiple-star systems. In addition, for the ZLK effect in triple systems, it is easier to trigger interactions between the inner binary (such as mergers and tidal dissipation) \citep{2016MNRAS.460.3494S,2016ARA&A..54..441N,2016ComAC...3....6T,2020A&A...640A..16T,2021MNRAS.502.4479H,2023ApJ...955L..14S,2024arXiv240706257S}. Simultaneously, many 3D simulations and observations of stellar mergers show that the merger products may differ from normally born stars \citep{2008A&A...488.1017G,2013MNRAS.434.3497G,2022MNRAS.516.1072C,2023MNRAS.519.5191B,2024arXiv240614416H}. Currently, the main method to identify merger products is asteroseismology (for example, by measuring the period spacing of gravity modes, asteroseismic mass, and the internal structure of merger products) \citep{2021MNRAS.508.1618R,2024arXiv240414474R,2024arXiv240614416H}. Although the merger process remains unclear, it is widely believed that the merger products exhibit many unique properties, such as unusual chemical compositions \citep{2015A&A...576L..12C,2015MNRAS.451.2230M,2018MNRAS.473.2984I,2019MNRAS.487.4343H}, amplified surface magnetic fields \citep{2009MNRAS.400L..71F,2014MNRAS.437..675W,2019Natur.574..211S,2024arXiv240702566P}, and unusually high \citep{2001ApJ...548..323S,2005MNRAS.358..716S,2011IAUS..272..531D,2019ApJ...881...47L} or low \citep{2016MNRAS.457.2355S,2019Natur.574..211S,2020MNRAS.495.2796S,2022NatAs...6..480W} rotational speeds. Understanding how these potential special properties affect merger products can provide new insights into the entire population of massive stars \citep{2020MNRAS.491..440W,2022NatAs...6..480W,2024ApJ...969..160L}. Therefore, it is essential to discuss whether the products of inner binary mergers in triple star systems evolve differently from normal stars and whether they can form $Gaia$ BH1.

In this $Letter$, we discuss an alternative formation scenario for $Gaia$ BH1, in which a highly rotating and strongly magnetic post-merger star produced by the inner binary merger can undergo CHE without a CEE, and form $Gaia$ BH1. Section 2 explains how rotation and magnetic fields affect the structure and evolution of the merger products. Section 3 discusses the origin and evolution of the triples that are progenitors of these post-merger binaries (PMBs), as well as the formation of the final $Gaia$ BH1. Section 4 provides the conclusions.

\section{The evolution of the inner binary merger product}
\subsection{Modelling of inner binary mergers}
Based on \cite{2013MNRAS.434.3497G} and \cite{2016MNRAS.457.2355S}, for the merger of two main sequence (MS) stars, hydrogen can mix into the center of the merger product as fuel, rejuvenating the merger product. Therefore, following \cite{1997MNRAS.291..732T}, \cite{2002MNRAS.329..897H}, \cite{2021MNRAS.502.4479H}, and \cite{2024ApJ...969..160L}, we assume that when the inner binary merges during the MS phase, the merger product can be rejuvenated to a zero-age main sequence (ZAMS) star. Using MESA stellar evolution code \citep{2011ApJS..192....3P,2013ApJS..208....4P,2015ApJS..220...15P,2018ApJS..234...34P,2019ApJS..243...10P} (version 10398), we simulate the structure and evolution of the merger product from ZAMS to the end of carbon burning.  According to \cite{2023MNRAS.518.1057E} and \cite{2023MNRAS.521.4323E}, the [Fe/H] of $Gaia$ BH1 is approximately -0.2. This means its metallicity is about $0.6\sim0.7\ Z_{\odot}$. In the simulation, we use a metallicity of $Z$ = 0.7 $Z_{\odot}$ ($Z_{\odot}$ = 0.02) = 0.014 \citep{2012A&A...537A.146E}. We apply the Ledoux criterion and the standard mixing length theory ($\alpha_{\rm MLT} = 1.5$) for convection calculations \citep{1991A&A...252..669L}. The overshooting parameter is set to 0.335 \citep{2011A&A...530A.115B}, and the semi-convection parameter is 1 \citep{1991A&A...252..669L}. The stellar wind scheme is consistent with that of \cite{2011A&A...530A.115B}.

Although the mechanism for angular momentum transport during the merger process is unclear \citep{1999A&A...349..189S,2002A&A...381..923S,2019MNRAS.485.3661F,2020A&A...634L..16D}, simple assumptions, such as angular momentum conservation or off-center collisions, might lead to rapid rotation in the merger product \citep{2013ApJ...764..166D,2014ApJ...796...37S,2022MNRAS.516.1072C}.  Simulations by \cite{2020ApJ...896...50C} also suggested that some of the rapidly rotating stars in Betelgeuse can be explained by stellar mergers. In addition, in the work of \cite{2014ApJ...782....7D} and \cite{2014ApJ...796...37S}, binary mergers are one important channel to explain Be/ Oe stars. Be/ Oe stars are typical fast-rotating stars, and their rotation speeds are usually about $\ge80\%$ of the break-up velocities \citep{1982ApJS...50...55S,1998A&A...338..505N,2001A&A...368..912Y,2003A&A...411..229R}. Therefore, to quantify the effect of rotation, we use the ratio of the rotational angular velocity to the critical rotational angular velocity ($\omega/\omega_{\rm crit}$) (where $\omega_{\text {crit }}^{2}=\left(1-L / L_{\mathrm{Edd}}\right) G M / R^{3}$, here, $G$ is the gravitational constant. $L$, $M$ and $R$ are the luminosity, mass and radius of the merger product. $L_{\rm Edd}$ is the Eddington luminosity \citep{2013ApJS..208....4P}), with a value of $\omega/\omega_{\rm crit}$ = 0.4 or 0.8 in the simulation. The efficiency parameter for rotational mixing is set to 1/30 \citep{2011A&A...530A.115B}, and the inhibition parameter for chemical gradients is set to 0.1 \citep{2011A&A...530A.115B}. In addition, the rotation can increase the mass-loss rate by a factor of $\left(\frac{1}{1-\omega / \omega_{\text {crit }}}\right)^{\beta}$, where $\beta$ = 0.43 \citep{1998A&A...329..551L}.

Simultaneously, observations show that about 7\% $\sim$ 10\% of MS stars have strong magnetic fields \citep{2009ARA&A..47..333D,2015SSRv..191...77F,2015A&A...582A..45F}. Following \cite{2009MNRAS.400L..71F}, \cite{2014MNRAS.437..675W}, \cite{2019Natur.574..211S} and \cite{2024arXiv240702566P}, we assume that a dynamo during the merger process can generate strong magnetic fields through differential rotation. However, direct modeling of magnetic field generation and amplification during the merging process usually requires 3D magneto-hydrodynamic simulations, which are beyond the scope of this $Letter$. For simplicity, we use the SURF scheme proposed by \cite{2020MNRAS.493..518K} and \cite{2022MNRAS.517.2028K} to calculate the impact of magnetic fields on the merger product. Based on observations of some MS stars with strong magnetic fields, their magnetic fields can reach several thousand to tens of thousands of Gauss \citep{2013MNRAS.429..398P,2018MNRAS.475.5144S}. Therefore, in the simulations, we assume magnetic fields of 0 and 5000 Gauss to quantify their effects on the merger products. In the magnetic field scheme of \cite{2017MNRAS.466.1052P}, \cite{2020MNRAS.493..518K}, and \cite{2022MNRAS.517.2028K}, the magnetic field can influence the evolution of massive stars through mass-loss quenching and magnetic braking. Specifically, the mass-loss rate of the merger products with a strong magnetic field is reduced by the mass-loss quenching parameter $f_{\rm B}$, which is described as follows: \citep{2008MNRAS.385...97U,2009MNRAS.392.1022U,2017MNRAS.466.1052P,2019MNRAS.485.5843K,2020MNRAS.493..518K,2022MNRAS.517.2028K}
\begin{equation}
f_{B} = \left\{
\begin{array}{ll}
\frac{\dot{M}}{\dot{M_{B=0}}} = 1 - \sqrt{1 - \frac{1}{R_{c}}} & \text{if } R_{\mathrm{A}} < R_{K} \\
\frac{\dot{M}}{\dot{M_{B=0}}} = 2 - \sqrt{1 - \frac{1}{R_{c}}} - \sqrt{1 - \frac{0.5}{R_{K}}} & \text{if } R_{\mathrm{A}} > R_{K}
\end{array}
\right.
\end{equation}
Here, $\dot{M}$ is the mass-loss rate of the merger product, $\dot{M_{B=0}}$ is the mass-loss rate when $B$ = 0, $R_{\rm A}$ is the Alfv${\rm \acute{e}}$n radius, $R_{\rm K}$ is the Kepler corotation radius, and $R_{\rm c}$ is the maximum distance from the last closed magnetic field line to the merger product's surface, which is approximated as $R_{\star}$ + 0.7($R_{\rm A}$ - $R_{\star}$), where $R_{\star}$ is the radius of the merger product. In addition, magnetic braking can consume the total angular momentum of the merger product, thereby slowing its rotation speed. Specifically, the total angular momentum lost per unit time $\frac{d J_{B}}{d t}=\int_{M_{t}}^{M_{\star}} \frac{d j_{B}(m)}{d t} d m=\sum_{k=1}^{x} \frac{d j_{B, k}}{d t} \Delta m_{k}$, where, $M_{\rm t}$ is the mass between the internal layers from index $k$ to $x$; $M_{\star}$ is the mass of the merger product; $\Delta m_{\rm k}$ is the mass of a given layer; $k$ is the number of layers from the surface inward for the merger product and $x$ is the number of layers in the last region not affected by mass-loss, typically on the order of 500 \citep{2020MNRAS.493..518K}.

Therefore, based on these observational facts and theoretical simulations, we will focus on the effects of rotation and magnetic fields on merger products in the next section.

\subsection{The effects of rotation and magnetic field on stellar evolution}
\begin{figure*}[htb]
\centering
\includegraphics[width=\textwidth]{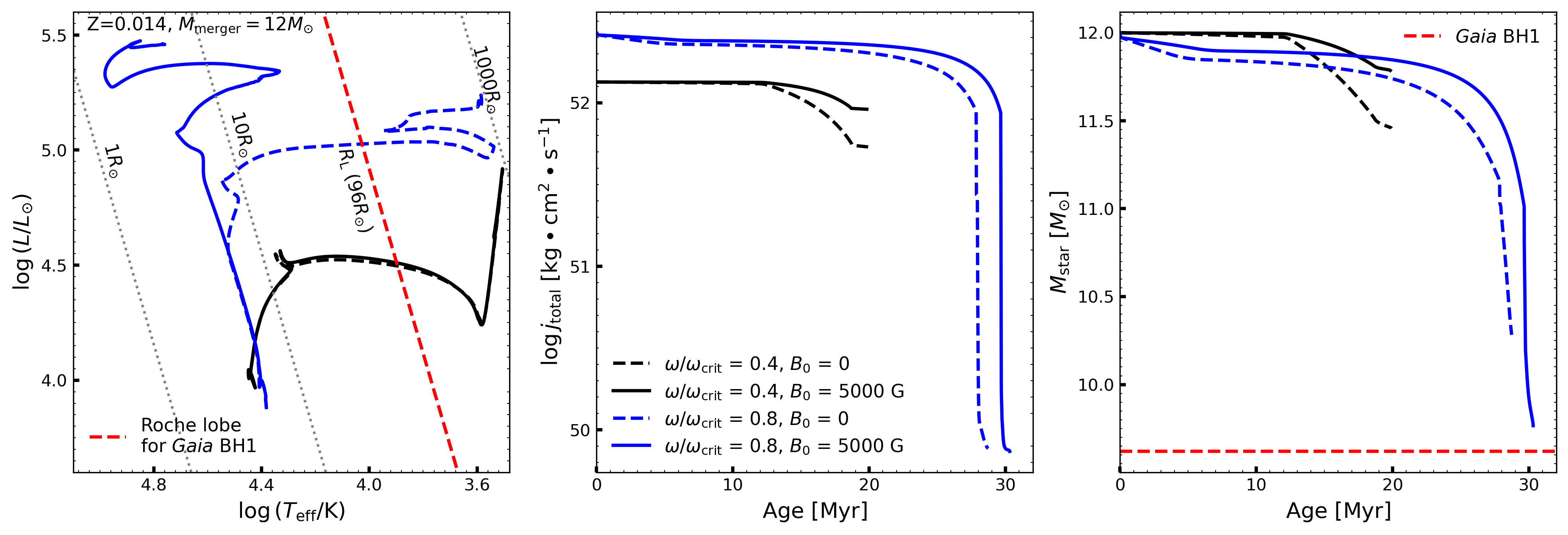}
\caption{We use the MESA code to calculate the evolutionary tracks of the 12 $M_{\odot}$ merged products at $Z$ = 0.014, with different rotations ($\omega/\omega_{\rm crit}$ = 0.4 and 0.8) and different magnetic fields ($B_{\rm 0}$ = 0 and 5000 G). The left panel shows the evolutionary tracks in H-R diagram. The black dotted line and the red dashed line represent the stellar equivalent radius lines and the Roche lobe radius of the observed $Gaia$ BH1 at perihelion. The middle and right panels show the total angular momentum and total mass of the merged products as a function of time, respectively.}
\label{fig:1}
\end{figure*}

Based on the properties of $Gaia$ BH1, the progenitors of $Gaia$ BH1 should prevent filling their Roche lobes, thereby to avoid CEE. We use the MESA code to calculate the evolutionary tracks of the merger product under different conditions. As shown in Fig. \ref{fig:1}, rotation and magnetic fields significantly affect the structure and evolution of merger products. Following \cite{1983ApJ...268..368E}, we calculate the Roche lobe radius of the current $Gaia$ BH1 at perihelion ($\sim$96 $R_{\odot}$), which is shown by a red dashed line in the left panel of Fig. \ref{fig:1}. In Fig. \ref{fig:1}, when considering magnetic field mass-loss quenching, magnetic braking, rotation, and stellar winds, we find that strong magnetic fields allow rapidly rotating merger products to retain more angular momentum for continued evolution (see the middle panel of Fig. \ref{fig:1}). This result is very similar to the conclusions of \cite{2020MNRAS.493..518K} and \cite{2023ApJ...952...79L}. This means that merger products with rapid rotation and strong magnetic fields have thinner H-rich envelopes, larger core masses, and smaller radii. Consequently, they naturally evolve towards the blue region of the H-R diagram. Typically, for a 12 $M_{\odot}$ non-magnetic, rapidly rotating merger product, although it initially trends towards the blue region of the H-R diagram, the higher mass-loss rate causes it to lose a significant amount of rotational angular momentum during subsequent evolution, leading to a rapid decrease in its rotation speed. This merger product undergoes a brief and inefficient CHE, causing its radius to significantly expand during subsequent evolution, eventually exceeding the Roche lobe radius of the current $Gaia$ BH1. It is therefore unlikely to be the progenitor of $Gaia$ BH1. In our simulations, compared to $\omega/\omega_{\rm crit}$ = 0.8, the merger product with $\omega/\omega_{\rm crit}$ = 0.4 has a lower rotation speed and less angular momentum. Simultaneously, significant mass loss and magnetic braking also take away much of its angular momentum, resulting in the merger product with $\omega/\omega_{\rm crit}$ = 0.4 not being able to undergo effective CHE. In the left panel of Fig. \ref{fig:1}, our model shows that merger products with rapid rotation ($\omega/\omega_{\rm crit} = 0.8$) and a strong magnetic field ($B_{\rm 0} = 5000 \ G$) can evolve through CHE into a helium star with a very small radius. Throughout the entire evolution, their maximum radius only reaches 30 $R_{\odot}$$\sim$50 $R_{\odot}$ and does not fill the Roche lobe radius of $Gaia$ BH1. As a comparison, the radius of non-magnetic star and lower-rotation stars ($\omega/\omega_{\rm crit}$ = 0.4) can reach up to hundreds of $R_\odot$.

On the other hand, The separation and eccentricity of $Gaia$ BH1 suggest that a low natal kick velocities ($V_{\rm kick}$) ($V_{\rm kick}$ $<$ 40$\sim$45 ${\rm km/s}$ \citep{2023MNRAS.518.1057E,2024arXiv240313579K}) was likely present, assuming a binary/ triple formation channel. This means that the progenitor of $Gaia$ BH1 may not have a significant amount of ejected mass during SN \citep{2012ApJ...749...91F}. In the right panel of Fig. \ref{fig:1}, our model shows that a merging product with an initial mass of $\sim$12 $M_{\odot}$, rapid rotation, and a strong magnetic field can form a BH with a mass of about 9.6 $M_{\odot}$ (red dashed line) with a relatively low ejected mass ($\sim$0.2 $M_{\odot}$). This is very close to the observed mass of $Gaia$ BH1.

\section{The formation of $Gaia$ BH1 through the merger of inner binary within the triples}
\subsection{Modelling the evolution of the triple}
If binary mergers can form rapidly rotating and strongly magnetic stars \citep{2001ApJ...548..323S,2005MNRAS.358..716S,2005AJ....129..466C,2019ApJ...881...47L},
it is crucial to identify and study the evolutionary paths that create the progenitor triples of $Gaia$ BH1 \citep{2019ApJ...881...47L,2024arXiv240706257S}. For the evolution of triples, only when the triple is initially dynamically stable and hierarchical can differ from isolated binary evolution models. This is because for triples that are initially unstable, the system will either merge or eject stars on a very short timescale, or evolve chaotically \citep{2024A&A...684A..35B}. When searching for progenitor triples that can form $Gaia$ BH1, we first use the analytical formula from \cite{2001MNRAS.321..398M} to determine if the initial triple is stable and hierarchical, i.e.,
\begin{equation}
\frac{a_{\rm out}}{a_{\rm in}}>2.8\left(1+\frac{m_{3}}{m_{1}+m_{2}}\right)^{\frac{2}{5}} \frac{\left(1+e_{\rm out}\right)^{\frac{2}{5}}}{\left(1-e_{\rm out}\right)^{\frac{6}{5}}}\left(1-\frac{0.3 i}{180^{\circ}}\right)
\end{equation}
Here, the subscripts "in" and "out" denote the inner and outer parts of the triple, respectively. Subscripts 1, 2, and 3 refer to the primary, the secondary in the inner binary, and third component, respectively. The symbol $i$ represents the mutual inclination between the pairs of orbits. Second, we also require that the two inner stars do not fill their Roche lobes at periastron on the ZAMS. Finally, we exclude triples that become unstable within a few Lyapunov timescales. Specifically, we use the publicly available "Ghost Orbits" code provided by \cite{2022MNRAS.516.4146V,2023MNRAS.525.2388V} to calculate the instability probability of the initial triple system within 100 outer orbit periods. We only consider initial triples with an instability probability $P\ <\ 10^{-4}$ \citep{2022MNRAS.516.4146V,2023MNRAS.525.2388V}. For the initial triples that meet all the above criteria, we use the Multiple Stellar Evolution (MSE) codes \citep{2021MNRAS.502.4479H} to simulate their evolution. The MSE code implements stellar evolution, binary evolution (including the effects of mass transfer in eccentric binaries and CEE, etc.), and the secular dynamical evolution of multiple-star systems. Additionally, the MSE code switches between direct N-body integration and secular equations of motion based on the system's stability at each integration step. In other words, the MSE code can simulate the evolution of multi-star systems through both unstable and stable phases \citep{2021MNRAS.502.4479H}.

The metallicity for all triples is set to $Z$ = 0.014. Note that any parameters not mentioned in the MSE code follow the default values described in \cite{2021MNRAS.502.4479H}. In the MSE code calculations, when the inner binary merges, it generates a recoil velocity that may cause the third component to become unbound \citep{2021MNRAS.502.4479H}. We only select PMBs that remain bound after the inner binary merges. At this point, we use the binary module of the MESA code to evolve the PMB until the merger product reaches core carbon depletion. As described in Section 2, we assume that the merger product has $\omega/\omega_{\rm crit} = 0.8$ and $B_{\rm 0} = 5000$ G.

Finally, we perform semi-analytical simulations of SN and FSN for the PMB to determine if it can form $Gaia$ BH1. Following \cite{2011ApJ...730...70O}, \cite{2013ApJ...769..109L}, and \cite{2023ApJ...952...79L}, CHE stars are likely to undergo a FSN and collapse directly into a BH. However, we do not exclude the possibility that merger products could undergo a SN. According to the study by \cite{2011ApJ...730...70O} and \cite{2013ApJ...769..109L}, stars that experience FSN typically receive very low $V_{\rm kick}$. We establish Model A and Model B to respectively address the cases where merger products undergo SN or FSN: (Model A: SN) The merger product undergoes a core-collapse supernova (CCSN). For the BH's received $V_{\rm kick}$, it is sampled from a Maxwellian distribution with a dispersion ($\sigma_{\rm k}$) of 50 ${\rm km/s}$ \citep{2006ApJ...644.1063D,2019ApJ...885..151S}. (Model B: FSN) The merger product experiences a FSN. We adopt the assumption from \cite{2011ApJ...730...70O} and \cite{2013ApJ...769..109L} that the merger product has a very low $V_{\rm kick}$, with $V_{\rm kick}$ sampled from a Maxwell distribution with a $\sigma_{\rm k}$ = 10 ${\rm km/s}$.

\subsection{Formation of $Gaia$ BH1}
The formation process of $Gaia$ BH1 is simulated using a stellar model that includes triple evolution, as well as a binary model with rapid rotation and a strong magnetic field. Given the properties of $Gaia$ BH1 and Equation (2), this means that the $P_{\rm out}$ of the progenitor triple of $Gaia$ BH1 cannot be too distant (about 100$\sim$130 days) and $P_{\rm in}$ must be very small. Additionally, in the PMB model of this $Letter$, the inner binary merger product cannot fill its Roche lobe, and there is no significant mass transfer. The masses, spins and orbital changes, including variations in the orbital period and eccentricity are primarily caused by stellar winds and/ or tidal interactions \citep{2002MNRAS.329..897H}.

We uniformly randomly generate initial input parameters through Monte Carlo simulation, which are input to the MSE code for computation until a $Gaia$ BH1-like orbit is formed, provided that all the conditions of the previous section are met. In our model, the initial outer orbital separation ($a_{\rm out}$), the outer orbital eccentricity ($e_{\rm out}$), and the total mass of the inner binary ($M_{1}+M_{2}$) are the most sensitive parameters. The reason is that the orbital properties of PMB and the masses of its components largely depend on these initial parameters. We selected an initially stable hierarchical triple that evolves into a PMB and eventually forms $Gaia$ BH1. The initial masses of the three stars in the triple system are 9.03 $M_\odot$, 3.12 $M_\odot$, and 1.00 $M_\odot$; the inner and outer orbital periods are 2.21 days and 121.92 days, respectively; and the inner and outer eccentricities are 0.41 and 0.45, respectively. Additionally, the inner and outer inclinations (radians), which are 1.47 and 2.96, respectively; the inner and outer arguments of pericentre (radians), which are 4.77 and 4.23, respectively; and the inner and outer longitudes of ascending node (radians), which are 0.81 and 6.13, respectively.

\begin{figure*}[htb]
\centering
\includegraphics[width=\textwidth, height=1\textheight]{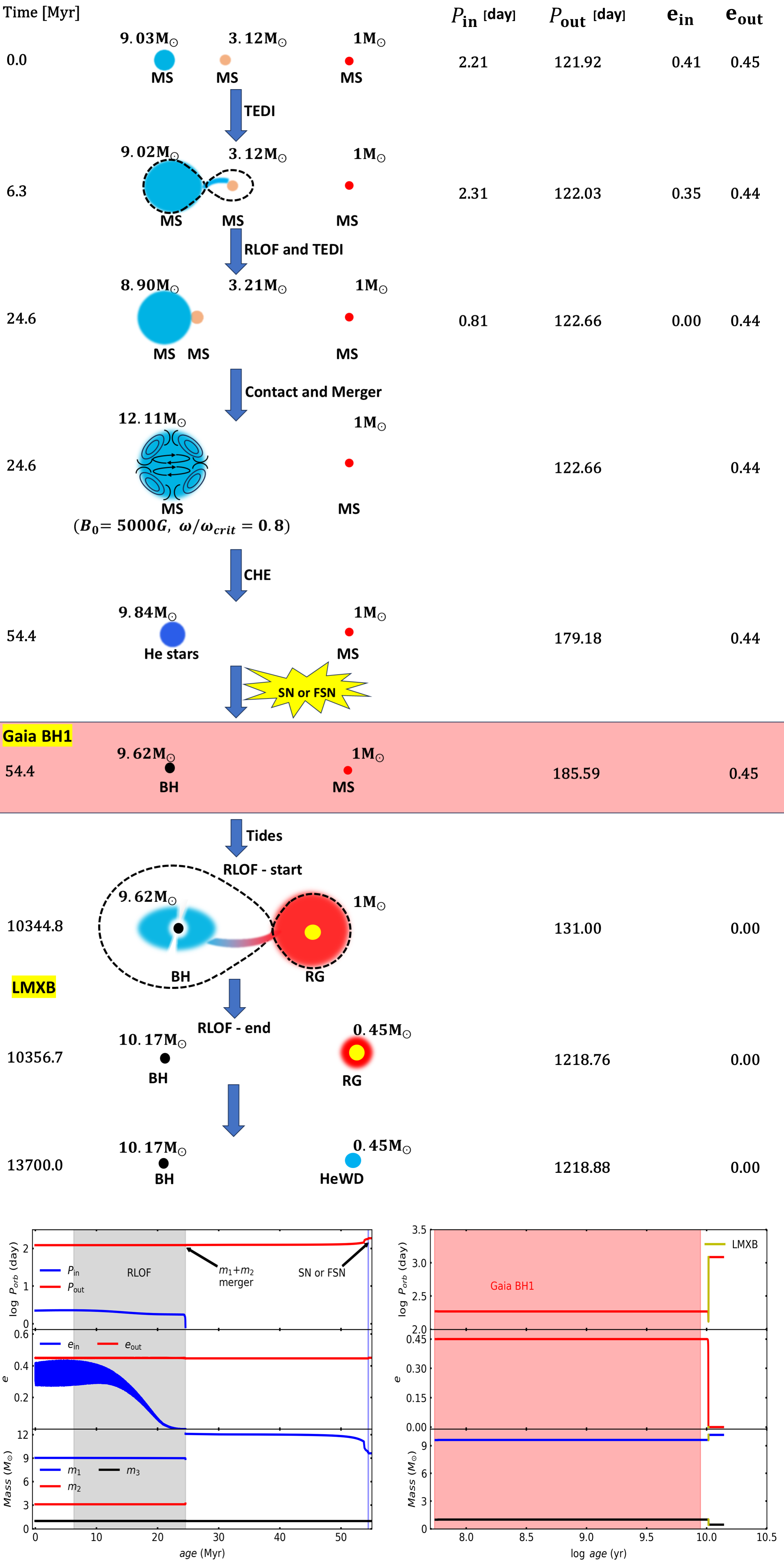}
\caption{The upper part shows the entire formation process from the triple to PMB to $Gaia$ BH1. The lower part shows the orbit period, eccentricity and mass as a function of time. The gray and red areas represent the RLOF phase and the GaiaBH1 phase, respectively.}
\label{fig:2}
\end{figure*}

Fig. \ref{fig:2} shows most of physical parameters for the evolution of the selected triple. In this triple system, because the masses of the inner binary are high, the inner binary evolves first. The stellar wind causes the inner and outer orbits to expand and creates the triple evolution dynamical instability (TEDI). This is consistent with the conclusion of \cite{2016ARA&A..54..441N}, \cite{2022MNRAS.516.1406S}, \cite{2022ApJ...925..178H}, and \cite{2024ApJ...964...83G}. In addition, the ZLK effect caused by angular momentum exchange between the inner and outer orbits leads to oscillations in their eccentricities during secular evolution \citep{2016ARA&A..54..441N}. $e_{\rm in}$ oscillates between about 0.29 and 0.43, while $e_{\rm out}$ oscillates between about 0.44 and 0.45. On the other hand, another main reason for TEDI in Fig. \ref{fig:2} is that the timescale of angular momentum changes due to long-term evolution is shorter than the orbital timescale. Specifically, it is described by the following equation \citep{2012ApJ...757...27A,2014ApJ...781...45A,2016MNRAS.458.3060L,2018MNRAS.481.4907G,2018MNRAS.481.4602L,2019MNRAS.490.4756L,2020MNRAS.494..850H}:
\begin{equation}
t_{j_{k}} \equiv\left|\frac{1}{j_{k}}\left(\frac{d j_{k}}{d t}\right)_{\mathrm{sec}}\right|^{-1}<P_{\mathrm{orb}, k}
\end{equation}
where $k$ describes any orbit, and $j_{\rm k}$ represents the specific angular momentum of that orbit. When the triple evolves to $\sim$6.3 Myr, the primary in the inner binary fills its Roche lobe and starts to experience Roche lobe overflow (RLOF). This means the inner orbit begins to shrink, tidal effects increase, and $e_{\rm in}$ starts to circularize. According to \cite{1997Natur.386..254H}, \cite{2001ApJ...562.1012E}, \cite{2002ApJ...578..775B}, \cite{2017MNRAS.467.3066A} and \cite{2024MNRAS.527.9782D}, when the timescale of tidal forces is shorter than the precession caused by triple dynamics, the ZLK oscillation effect will gradually be quenched by the short range tidal force. Therefore, in the selected triple, the oscillation amplitudes of $e_{\rm in}$ and $e_{\rm out}$ gradually decrease. After about 25 Myr, $P_{\rm in}$ of the triple reduces to 0.81 days, $e_{\rm in}$ is circularized to 0, and the companion's radius fills its Roche lobe, satisfying the condition $\left(R_{1}+R_{2}\right)>a_{i n}\left(1-e_{i n}\right)$. In the MSE code, such a system is treated as if it has merged, which is roughly analogous to the secondary in a contact binary overflowing its second Lagrangian point \citep{2002MNRAS.329..897H,2017PASA...34...58E,2018MNRAS.474.2959G,2021MNRAS.502.4479H,2022ApJS..258...34R}. Such binaries are typically treated as merged \citep{2021MNRAS.507.5013M,2022A&A...667A..58P,2024ApJ...969..160L}.

In the time series diagram of Fig. \ref{fig:2}, after the inner binary merges, PMB consists of a 12.11 $M_{\odot}$ star and a 1 $M_{\odot}$ companion, with an orbital period of 122.66 days and an eccentricity of 0.44. Based on our assumptions, the star produced via binary merger has $\omega/\omega_{\rm crit} = 0.8$ and $B_{\rm 0}$ = 5000 G. Due to the angular momentum loss resulting from the mass loss of the merged product, the orbital period of PMB becomes wider. Additionally, because PMB has a wide orbital separation and the radius of the merged product is much smaller than its Roche lobe radius, orbital circularization is unlikely. At about 54 Myr, the merged product forms a 9.84 $M_{\odot}$ helium star through CHE and reaches core carbon depletion. At this point, the orbital period of PMB is 179.18 days, with an eccentricity of 0.44.

\begin{figure*}[htb]
\centering
\includegraphics[width=\textwidth]{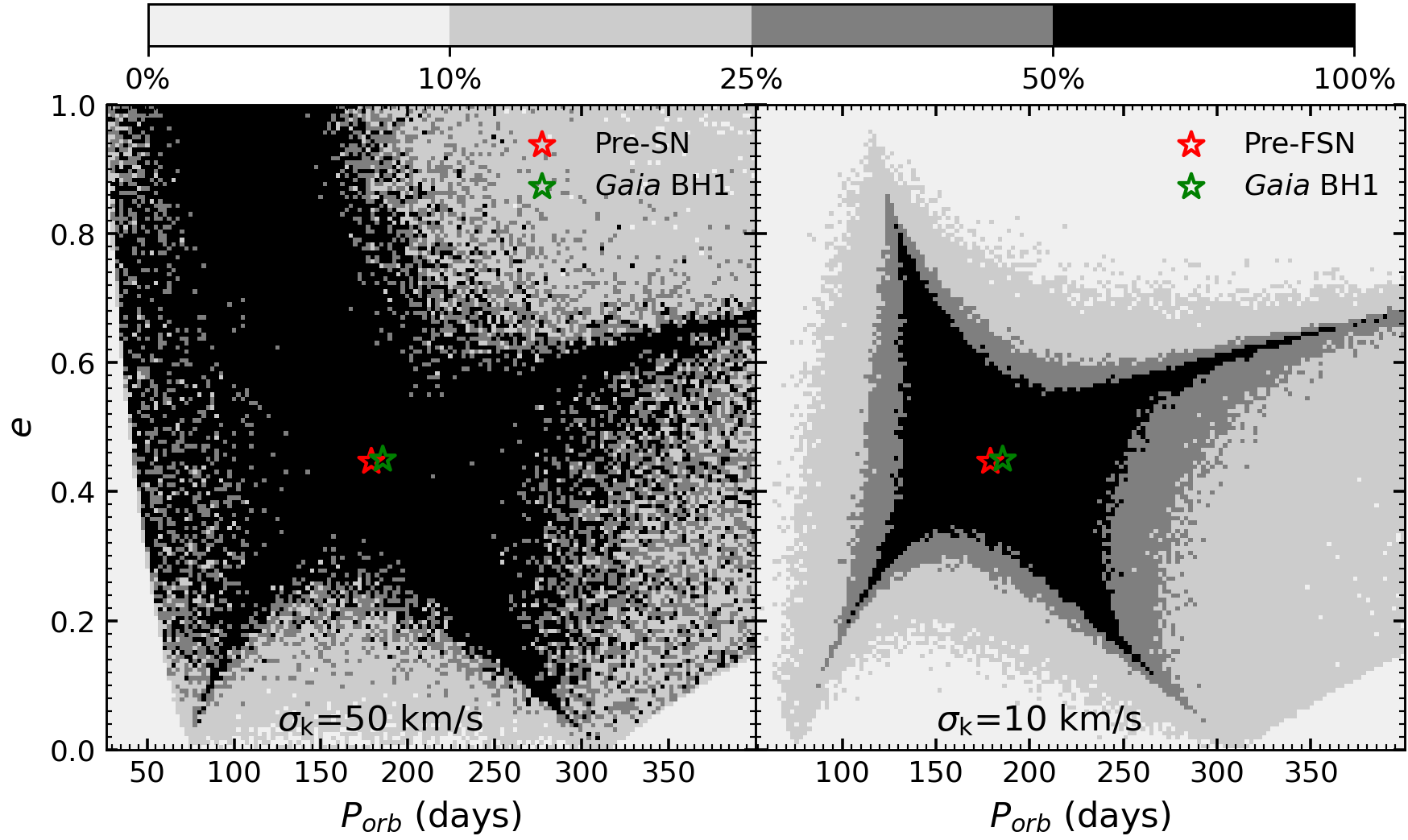}
\caption{The distribution of orbital periods and eccentricity for the core carbon depleted PMB simulations in Fig. \ref{fig:2}, after undergoing a SN (left panel) or a FSN (right panel). The red and green stars represent the positions of the PMB before the SN or FSN and $Gaia$ BH1, respectively. The color mapping represents the density of a certain area; the darker the color, the more samples there are in that region, indicating higher density.}
\label{fig:3}
\end{figure*}

Following \cite{2022A&A...667A..58P}, \cite{2023A&A...674A.216L}, and \cite{2023ApJS..264...45F}, we assume that the merger product will undergo the core collapse after core carbon depletion, resulting in a SN or a FSN. In the simulation of Fig. \ref{fig:2}, regardless of whether $Gaia$ BH1 forms through a SN or a FSN, the ejected mass requires $\sim$ 0.22 $M_{\odot}$. Following \cite{2012ApJ...747..111W}, \cite{2013ApJ...769..109L} and \cite{2023A&A...673L..10A}, the energy imparted to the merged product by a FSN ($\sigma_{\rm k}$ $\sim$ 10 ${\rm km/s}$) is typically lower than that from a SN ($\sigma_{\rm k}$ $\sim$ 50 ${\rm km/s}$). Based on the Models A and B described in the previous section, we perform SN and FSN simulations for this PMB system, respectively, and the results are shown in Fig. \ref{fig:3}. The PMB system in FSN simulations shows less dispersion, generally experiences smaller natal kicks compared to those in SN simulations, and has a higher survival rate of 99\%, whereas the survival rate in SN simulations is about 51\%. In addition, the energy received by the remnant in the SN model is typically larger ($\sigma_{\rm k}$ is larger) than that received in the FSN model \citep{2011ApJ...730...70O,2013ApJ...769..109L}, which eventually leads to BH receiving a generally larger $V_{\rm kick}$ in the SN model. In our model, the $V_{\rm kick}$ for $Gaia$ BH1 formed through a SN or a FSN is about $46^{+25}_{-33}$ ${\rm km/s}$ or $9^{+16}_{-8}$ ${\rm km/s}$. By comparing the orbital properties of SN and FSN before and after the event, as well as those observed for $Gaia$ BH1 in Fig. \ref{fig:3}, we find that the observational data of $Gaia$ BH1 is primarily located in regions of higher density (probability) after a SN or a FSN (black region).

Finally, we also use the MESA code to calculate the subsequent evolution of the formed $Gaia$ BH1 (see Fig. \ref{fig:2}). At $\sim$10.34 Gyr, the secondary (the previous  tertiary) evolves into a red giants (RGs), and its radius rapidly increases so that it fills its Roche lobe. Then, mass transfer (MT) occurs, and the quiescent BH binary turns into a low-mass X-ray binary (LMXB). In the process of MT, the orbital separation gradually widens and the Roche lobe radius increases because the donor is less mass than the gainer. When the Roche lobe radius becomes larger than the donor's radius, MT stops. At this point, the orbital period is 1218.76 days, with $M_{\rm BH}$ = 10.17 $M_{\odot}$ and $M_{\rm RG}$ = 0.45 $M_{\odot}$, and the LMXB stage lasts $\sim$11.9 Myr. In the subsequent evolution, this RG loses its remaining hydrogen envelope ($\sim10^{-4}$ $M_{\odot}$) through stellar wind and eventually cools to form a 0.45 $M_{\odot}$ helium white dwarf (HeWD).

\section{Conclusion and discussion}
Following the analysis of \cite{2023MNRAS.521.4323E}, $Gaia$ BH1 is difficult to explain through isolated binary evolution models or requires a very large $\alpha_{\rm CE}$ to form. Therefore, in this $Letter$, we investigate the formation of $Gaia$ BH1 originates from a triple system, where the inner binary merges to form $Gaia$ BH1. Based on the high metallicity environment of $Gaia$ BH1, we first use the MESA code to study the structure and evolution of stars with a metallicity of $Z$ = 0.014, rapid rotation, and different magnetic field strengths. We find that if a massive star initially has rapid rotation and a strong magnetic field (here we assume $\omega/\omega_{\rm crit} = 0.8$ and $B_{\rm 0} = 5000$ G), it can retain a large amount of angular momentum during its evolution and undergo CHE. Throughout its evolution, its radius is much smaller than that of a non-magnetic massive star and does not fill the Roche lobe radius of $Gaia$ BH1.

Subsequently, we validate and search for potential progenitor triple systems that form $Gaia$ BH1. Using Equation 2 and the code provided by \cite{2022MNRAS.516.4146V} and \cite{2023MNRAS.525.2388V}, we ensure that the initially generated triple is hierarchical and stable. Then, through calculations with the MSE and MESA codes, we identifie and evolve a hierarchical triple that can form $Gaia$ BH1. This triple first goes through TEDI and then through RLOF, after which the inner binary merged through contact (or collision), forming a PMB. Under the assumptions of rapid rotation and strong magnetic field, a PMB consisting of a merger product with a mass of 12.11 $M_{\odot}$ and a companion (originally an outer tertiary) with a mass of 1 $M_{\odot}$ could, after undergoing CHE, ultimately form $Gaia$ BH1 in the Galactic disk via a SN or a FSN.

However, our model still has many limitations and uncertainties. First, whether the merger product is rotating and has a strong magnetic field still requires extensive simulations and more observational data to obtain better constraints and clear conclusions. Secondly, understanding the evolution of merger products with rapid rotation and/ or strong magnetic fields and triples may require higher dimensional modelling (e.g. 3D) and numerical studies. Finally, we only test one set of initial conditions, which is clearly insufficient for understanding more BH-like systems.

On the other hand, our model is expected to explain the $Gaia$ BH2, BH3 and $Gaia$ BH-like systems. The reason is that our model does not involve CEE, which avoids the risks of insufficient orbital energy and/or merging. This is a unique feature of this channel. Therefore, in contrast to the work of \cite{2024ApJ...964...83G} explaining $Gaia$ BH1, our model is expected to be insensitive to the $\alpha_{\rm ce}$ parameter. In addition, consider the fact that a very high fraction of massive star is born in triple systems \citep{2014AJ....147...86T,2017ApJS..230...15M,2023ASPC..534..275O}, while about 35$\sim$40\% of triple systems experience inner binary mergers \citep{2024arXiv240811128B}. Therefore, this implies that the rate (or number) of $Gaia$ BH-like systems formed by our model may be non-negligible. However, these insights still require future calculations via detailed population synthesis methods to reach more definitive conclusions.

\section*{Acknowledgements}
This work received the support of the National Natural Science Foundation of China under grants U2031204, 12163005, 12373038, and 12288102; the Natural Science Foundation of Xinjiang No.2022D01D85 and 2022TSYCLJ0006; the science research grants from the China Manned Space Project with No.CMS-CSST-2021-A10.

\bibliography{./sample631}
\bibliographystyle{./aasjournal}
\end{document}